\documentclass[prl,aps,twocolumn,showpacs]{revtex4}
\usepackage{graphicx}
\begin{document}

\title{Quantum phases in a resonantly-interacting Bose-Fermi mixture}
\author{D. B. M. Dickerscheid,$^{1,2}$
        D. van Oosten,$^{1,3}$
        E. J. Tillema,$^1$ and
        H. T. C. Stoof$^1$}
\affiliation{$^1$Institute for Theoretical Physics, Utrecht
University, Leuvenlaan 4, 3584 CE Utrecht, The Netherlands}
\affiliation{$^2$Lorentz Institute, Leiden University, P.O. Box
9506, 2300 RA Leiden, The Netherlands}
\affiliation{$^3$Johannes
Gutenberg Universit\"at, 55099 Mainz, Germany}

\date{\today}

\begin{abstract}
We consider a resonantly-interacting Bose-Fermi mixture of
$^{40}$K and $^{87}$Rb atoms in an optical lattice. We show that
by using a red-detuned optical lattice the mixture can be
accurately described by a generalized Hubbard model for $^{40}$K
and $^{87}$Rb atoms, and $^{40}$K-$^{87}$Rb molecules. The
microscopic parameters of this model are fully determined by the
details of the optical lattice and the interspecies Feshbach
resonance in the absence of the lattice. We predict a quantum
phase transition to occur in this system already at low atomic filling 
fraction, and present the phase diagram as a function of the temperature and 
the applied magnetic field.
\end{abstract}
\maketitle

{\it Introduction.} --- In the last few years experiments have shown that it is
possible to realize a quantum degenerate gas of
fermionic atoms \cite{DeMarco}. Combining such a degenerate Fermi gas
with a Bose-Einstein condensate of atoms, it is also possible to obtain
a quantum degenerate Bose-Fermi mixture \cite{Truscott}, thus
creating a dilute analog of the liquid $^3$He-$^4$He mixture. The recent
observation of interspecies Feshbach resonances in a Bose-Fermi
mixture \cite{Inouye2004a,Stan2004a} opens up even richer physics,
as this couples the fermionic and bosonic atoms to a third
species, namely the fermionic heteronuclear molecules. An
interesting aspect of the resonantly-interacting mixture
that we address in much more detail in the following, is the
possibility to reversibly destroy a Bose-Einstein
condensate of atoms, by associating the bosonic atoms into
fermionic molecules and thus creating a degenerate Fermi gas of dipolar 
particles. 
Moreover, the Bose-Einstein condensed phase is very interesting by 
itself because it contains a macroscopic quantum coherence between the
fermionic atoms and molecules.
However, in experiments with magnetic or
optical traps the molecules quickly decay due to atom-molecule and
molecule-molecule collision \cite{MIT}. By loading the degenerate
Bose-Fermi mixture into an optical lattice with a total filling factor
less then unity, these collisions can be prevented and the
lifetime of the molecules is expected to be dramatically enhanced.

Bose-Fermi mixtures in an optical lattice, but in the absence of
an interspecies Feshbach resonance, have been the subject of
active theoretical investigation lately. 
In particular, lattice symmetry breaking \cite{Albus2003a}, the existence of 
quantum phases that involve the pairing of fermions with bosons
\cite{Lewenstein2004a}, and also unconventional fermion pairing
\cite{Demler2005a} have been predicted. It is the main objective
of this Letter to generalize these studies to the case of a
resonantly-interacting Bose-Fermi mixture. Although our
theoretical methods are very general, we consider as a concrete
example a mixture of fermionic $^{40}$K atoms
and the bosonic $^{87}$Rb atoms, as this system is now becoming
available experimentally \cite{Esslinger2004a}. A gas consisting of these two
atoms is especially promising as their wavelength is relatively
close and readily accessible using Ti:Sapphire lasers. Furthermore,
the mass of $^{40}$K is much larger than the mass of the other
experimentally available fermionic atom ${^6}$Li, which makes it
much easier to trap this species in an optical lattice.

\begin{figure}[ht]
\includegraphics[width=.95\columnwidth]{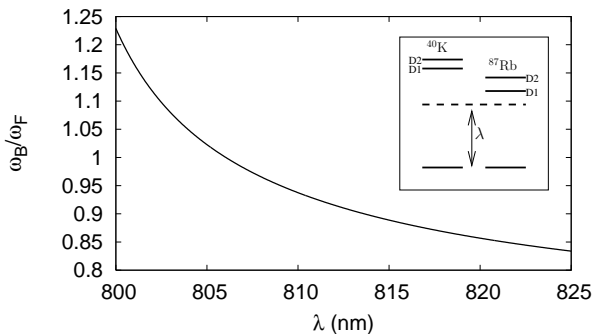}
\vspace{-.2cm}
\caption{Ratio of the on-site trap frequencies of the boson $^{87}$Rb
and the fermion $^{40}$K,
as a function of the lattice laser wavelength. The inset shows the
fine-structure levels of the two atomic species. }
\label{wavelengthpic}
\end{figure}

In order to analyze the properties of 
a resonantly-interacting Bose-Fermi mixture
most easily, it is convenient that all the species in the optical
lattice, i.e., the fermionic atoms, the bosonic atoms, and the
fermionic molecules, experience the same on-site trapping
frequency. Because the potassium atoms are lighter than the
rubidium atoms, the potential for the former should thus be less deep
than that for the latter. Moreover, as is shown in the inset of
Fig.~\ref{wavelengthpic}, both the D1 and D2 lines of potassium
are blue compared to the D1 and D2 lines of rubidium. As a result
equal on-site trapping frequencies can only be achieved in 
an optical lattice that is red detuned with respect to all
these four transitions.
Making use of the fact that the hyperfine structure of the atoms is no
longer resolved for the detunings used in optical
lattice experiments \cite{Grimm1999}, we have calculated the ratio of the
two on-site trapping frequencies as a function of the wavelength
of the lattice laser. As can be seen from
Fig.~\ref{wavelengthpic}, using a wavelength of $806$~nm, ensures
that the trapping frequencies for both atomic species are the
same. In principle the polarizability of the molecule is not
known. However, recent experiments have shown that for homonuclear
molecules, the resulting trap frequency for the molecules is
almost the same as that of the atoms \cite{Bloch2004}. 
In the following, we make the reasonable assumption
that this also holds for the heteronuclear $^{40}$K-$^{87}$Rb
molecules of interest to us.

For the particular detuning given above, the on-site trapping frequency 
is related to the Rabi frequency $\Omega$ of the lattice laser by
$\omega \equiv \omega_{F} = \omega_{B} = 2.1\cdot 10^{-5}\mbox{\,}\Omega$. Having to use
a red-detuned optical lattice has the disadvantage that the atoms are
trapped in the light and not in the dark, which in principle results in 
a larger decay
due to spontaneous emission.  We have therefore also calculated this emission
rate and find that $\Gamma/\Omega^{2} = 1.9\cdot 10^{-21} ~{\rm s}$ 
for the optimal wavelength.
Clearly the lifetime is always longer than one second for realistic Rabi 
frequencies less than $4$ GHz, and spontaneous emission can thus be safely
neglected.

{\it Generalized Hubbard model.} --- 
We now consider the many-body aspects of the Bose-Fermi mixture near a 
Feshbach resonance. We consider a mixture of
$^{40}$K and $^{87}$Rb that is loaded into a three-dimensional 
and cubic optical lattice
that is sufficiently deep, so that we are allowed to use a tight-binding
approximation for the band structure of the single-particle states. 
Experimentally this requires the Rabi frequency of the lattice laser
to be larger than about $1$~GHz. For the reasons mentioned
above, we consider only low filling fractions, which limits the
maximum number of atoms on a single site. As a result, we can neglect
atom-molecule and molecule-molecule interactions. Furthermore, for
low filling fractions there is also no problem of phase separation
of the atomic Bose-Fermi mixture \cite{Buchler2004a} and 
it is justified to neglect possible Mott physics in the mixture
\cite{Sengupta2004a,Holland2005a}. 

Under these conditions we have recently derived the theory for
resonantly-interacting ultracold atomic gases in an optical
lattice \cite{Dickerscheid2004a}. Within this theory the 
two-body Feshbach problem at a single site is solved
exactly, which physically leads to a dressing
of the molecules and to various avoided crossings in the on-site 
energy levels of two atoms $\epsilon_{\sigma} (B)$
as the magnetic field $B$ is swept through the Feshbach resonance. 
After having solved the on-site problem the various hopping parameters
can be calculated in the tight-binding approximation. In this manner
the microscopic
parameters of the generalized Hubbard model that describes the
Bose-Fermi mixture near the Feshbach resonance are completely
determined by the details of the optical lattice potential and the
experimentally known parameters of the Feshbach resonance in the
absence of the optical lattice. Ultimately, the mixture is
described by the following effective hamiltonian,
\begin{widetext}
\vspace{-.6cm}
\begin{eqnarray}\label{eq:bh}
H &=& -t_{\rm F} \sum_{\langle i, j \rangle } c^{\dagger}_{i}
c^{\phantom \dagger}_{j} -t_{\rm B} \sum_{\langle i, j \rangle }
a^{\dagger}_{i} a^{\phantom \dagger}_{j} -t_{\rm m} \sum_{\sigma} \sum_{\langle
i,j \rangle} b^{\dagger}_{i,\sigma} b^{\phantom
\dagger}_{j,\sigma} \nonumber + \sum_{i}  \left( \epsilon_{\rm a}
- \mu_{F} \right) c^{\dagger}_{i} c^{\phantom \dagger}_{i}
\nonumber \\ && + \sum_{i}  \left( \epsilon_{\rm a} - \mu_{B}
\right) a^{\dagger}_{i} a^{\phantom \dagger}_{i} +  \sum_{\sigma} \sum_{i}
\left( \epsilon_{\sigma} - \mu_{F} - \mu_B \right)
b^{\dagger}_{i,\sigma} b^{\phantom \dagger}_{i,\sigma} +
\frac{U^{BB}}{2} \sum_{i} a_{i}^{\dagger}a_{i}^{\dagger}
a_{i}^{\phantom \dagger}a_{i}^{\phantom \dagger} + U_{\rm
bg}^{BF} \sum_{i} a_{i}^{\dagger}c_{i}^{\dagger}
c_{i}^{\phantom \dagger}a_{i}^{\phantom \dagger} \nonumber \\
&& + g' \sum_{\sigma} \sum_{i} \sqrt{Z_{\sigma}} \left(
b^{\dagger}_{i, \sigma} c^{\phantom \dagger}_{i} a^{\phantom
\dagger}_{i} + a^{\dagger}_{i} c^{\dagger}_{i} b^{\phantom
\dagger}_{i,\sigma} \right).
\end{eqnarray}
\vspace{-.6cm}
\end{widetext}
Here $t_{\rm F}$, $t_{\rm B}$ and $t_{m}$ are the tunneling or hopping
amplitudes for the fermionic atoms, the bosonic atoms, and the
fermionic molecules, respectively. The symbol $\langle i,j \rangle$
denotes a sum over nearest neighbors. The operators
$c^{\dagger}_{i}$, $c^{\phantom \dagger}_{i}$ and
$a^{\dagger}_{i}$, $a^{\phantom \dagger}_{i}$ correspond to the
creation and annihilation operators of a single fermionic and
bosonic atom at site $i$, respectively. The operators
$b^{\dagger}_{i,\sigma}$, and $b^{\phantom \dagger}_{i,\sigma}$
correspond  to the creation and annihilation operators of the
various dressed molecules at site $i$. Also $\epsilon_{\rm
a} = 3 \hbar \omega/2$ is the on-site energy of a single atom. The
on-site interaction between two bosons is given by $U^{BB}$, and
$U_{\rm bg}^{BF}$ denotes the on-site background interaction
between the bosons and the fermions. In the tight-binding limit
the hopping amplitudes can be conveniently expressed in terms of
the lattice parameters as \cite{vanOosten2001a}
\begin{equation}
t_{\nu} = \frac{\hbar \omega}{2} \left( 1 - \left(\frac{2}{\pi} \right)^{2} \right)
\left( \frac{\lambda}{4 l_{\nu}} \right)^{2} e^{-(\lambda/4 l_{\nu})^{2} }.
\end{equation}
Here $\nu$ distinguishes the different species in the mixture,
i.e., $\nu = {\rm F}$ for the
fermionic atoms,  $\nu = {\rm B}$ for the bosonic atoms,
and $\nu = {\rm m}$ for the molecules.
Moreover, $\lambda$
is the wavelength of the light used to create the optical lattice,
and  the harmonic oscillator lengths obey
$l_{\nu} = \sqrt{\hbar / m_{\nu} \omega}$, 
where $m_{\rm F}$, $m_{\rm B}$, and $m_{\rm m} = m_{\rm F} + m_{\rm B}$
 are the masses of the fermions, bosons, and molecules, respectively. 
As expected, we have in general that $t_{\rm m} \ll
t_{F,B} \ll \hbar \omega$. 
Note that we have introduced a chemical potential for each atomic species,
since it is experimentally possible to control both the filling fraction of 
the fermions as well as the bosons in the mixture.

\begin{figure}
\includegraphics[angle=270,width=0.9\columnwidth]{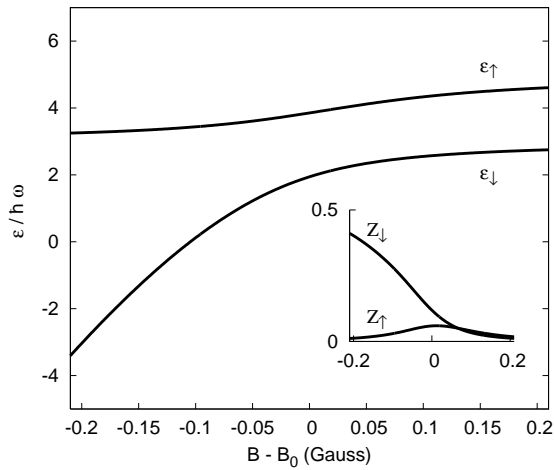}
\caption[Fig1]{
Details of the physical content of our theory.
We show the avoided crossing between the bare molecular level and the
lowest two-atom trap state.
The inset shows the probability $Z_{\sigma}$ as a function of
the magnetic field. This figure was
calculated for the $^{40}$K - $^{87}$Rb mixture with $\lambda = 806$~nm and
$\Omega = 1$ GHz.
}
\label{fig:avoided}
\end{figure}

Close to resonance we can neglect the on-site
interactions. The strength of the atom-molecule coupling in the
lattice is given by $g' = g/( \pi(l_{B}^{2} + l_{F}^{2}) )^{3/4}$, 
where $g = \sqrt{2 \pi a_{\rm bg} \Delta \mu \Delta B/ m_{\rm r}}$ 
is the bare atom-molecule coupling without the lattice
and $m_{\rm r} = m_{F} m_{B}/ (m_{F} + m_{B})$ is the reduced mass. 
For the  $^{40}$K-$^{87}$Rb mixture with potassium in the hyperfine state
$|9/2,-9/2\rangle$ and rubidium in the hyperfine state $|1,1\rangle$
there occurs a Feshbach resonance at $B_{0} = 510$~Gauss for which 
the  parameters that determine $g$ are given by 
the background scattering length $a_{\rm bg} = 150~a_{0}$
and the width of 
the resonance  $\Delta B = 1$~Gauss \cite{Inouye2004a,Thicknor}. 
The difference in magnetic moments $\Delta \mu$
 is equal to $29/22$ Bohr magneton in that case.
In Fig. \ref{fig:avoided} we show a
close-up of the avoided crossing and the wavefunction
renormalisation factors $Z_{\sigma}$ that give the probability 
for the dressed molecules to be in the bare molecular state of this
Feshbach resonance.
The probability $Z_{\sigma}$ is
determined by the selfenergy of the molecules $\hbar \Sigma_{\rm m} (E)
= (g^{2}/ \pi l_{r}^{3} \hbar \omega ) \cdot \Gamma(-E/2\hbar \omega + 3/4 )/\Gamma(-E/2\hbar \omega + 1/4 )$,
with $\Gamma(z)$ the gamma function, 
through the relation $Z_{\sigma} = 1/ (1 - \partial \hbar
\Sigma_{\rm m} (E)/\partial E )$ \cite{Dickerscheid2004a}. Note that in 
Fig.~\ref{fig:avoided} the sum of the probabilities $Z_{\downarrow} +
Z_{\uparrow}$ does not add up to one. This means that 
for this relatively broad interspecies Feshbach resonance a 
single-band approximation is not 
valid to determine the dressed molecular wavefunctions. However, a single-band
approximation in terms of dressed molecules is always possible for the low
filling fractions of interest to us, because the  higher-lying
on-site dressed molecular states will not be populated as we show now.

{\it Phase diagram.} ---
With the above formalism we next determine the phase diagram for the
$^{40}$K and $^{87}$Rb mixture. For simplicity we consider only equal densities
for both atomic species, although the generalization is immediate. 
We mentioned already that we consider a deep optical lattice 
for which the hopping strengths of the atoms are small
with respect to the level splitting in the on-site microtrap realized by the optical lattice.
The hopping strength of the molecules can be completely neglected in this limit and consequently
the band-structure for the molecules is essentially flat.
At zero temperature the energy $\epsilon_{\downarrow}$ of the
 lowest molecular state can,
depending on the magnetic field,
 be either smaller or larger than the sum of the lowest atomic 
energy levels. As a result, the ground state is either 
a Fermi sea of  $^{40}$K - $^{87}$Rb molecules, or a Fermi sea of $^{40}$K atoms and 
a Bose-Einstein condensate of $^{87}$Rb atoms.
Because of the different symmetries of these ground states there exists a 
quantum phase transition between these two states
that is in the same universality class as the XY model.
In the following, we calculate the phase diagram as a function of total 
filling fraction and temperature
by performing a mean-field analysis of the normal state of the 
hamiltonian in Eq. (\ref{eq:bh}).

For the equation of state for the total filling fraction
we find always that 
\begin{equation}\label{eq10}
n = n_{\rm F} + n_{\rm B} + 2 \sum_{\sigma} n_{{\rm m},\sigma}.
\end{equation}
In the normal state the molecular filling fractions obey
\begin{eqnarray}\label{eq11}
n_{{\rm m},\sigma} = \frac{1}{N_{s}} \sum_{{\bf k}}
\frac{1}{e^{\beta \hbar \omega_{{\bf k},\sigma}}+1},
\end{eqnarray}
where $\hbar \omega_{{\bf k},\sigma}= \epsilon_{{\bf k},\sigma} - (\mu_{F} + \mu_{B}) $
is the molecular dispersion.
The atomic filling fractions are given by
\begin{eqnarray}\label{eq12}
n_{\rm F} &=& \frac{1}{N_{s}} \sum_{{\bf k}}
\frac{1}{e^{\beta \hbar \omega_{{\bf k},F}} + 1},
\nonumber \\
 n_{\rm B} &=&  \frac{1}{N_{s}} \sum_{{\bf k}}
\frac{1}{e^{\beta \hbar \omega_{{\bf k},B}}-1},
\end{eqnarray}
where 
\mbox{ $\hbar \omega_{{\bf k},\rm F} = \epsilon_{{\bf k},\rm F} - \mu_{F}$}
and
\mbox{$\hbar \omega_{{\bf k},\rm B} = \epsilon_{{\bf k},\rm B} 
- \mu_{B}$}
are the dispersion relations for the fermionic and the bosonic atoms,
respectively. 
Here \mbox{$\epsilon_{{\bf k},\rm F,B} = -2t_{\rm F,B} 
\sum_{j=1}^{3} \cos{(k_{j} \lambda /2)} + \epsilon_{\rm a}$} for the atoms
and 
$\epsilon_{{\bf k},\sigma} = \epsilon_{\sigma}$
 for the molecules. 
\begin{figure}
\includegraphics[width=1.0\columnwidth]{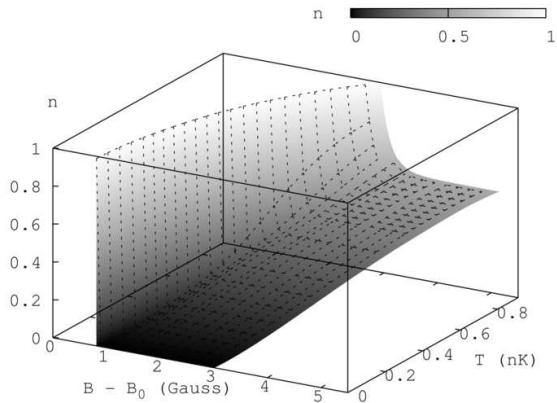}
\caption[Fig2]{
Phase diagram of the 
$^{40}$K and $^{87}$Rb mixture as a function of the total filling fraction,
magnetic field and temperature.
This figure was
calculated for the $^{40}$K - $^{87}$Rb mixture with $\lambda = 806$~nm and
$\Omega = 1$~GHz.
}\label{fig3.eps}
\end{figure}
We have seen that at zero temperature there is a quantum phase transition between a phase where the $^{87}$Rb atoms are Bose-Einstein condensed and a phase
with only a Fermi sea of molecules. The quantum critical point is
determined by the ideal gas condition for Bose-Einstein
condensation, i.e.,
 $\epsilon_{\downarrow} = 2 \epsilon_{\rm a} - 6 (t_{\rm B} + t_{\rm F})$. 
In the approximation that we can neglect the hopping of the molecules, their
band structure is flat and the  quantum critical point
is independent of the filling fraction of the molecules.
Including the molecular band structure would lead to a critical 
magnetic field that 
slowly shifts to lower magnetic fields as the density increases.
 At nonzero temperatures we can also determine the critical temperature 
as a function of the detuning and the total filling
fraction from the equation of state. The phase diagram in Fig.
\ref{fig3.eps} shows how at constant total atomic filling fraction 
the critical temperature depends on the magnetic field. For large enough magnetic fields
there are no molecules and the critical temperature is determined 
for low densities by the critical temperature of an ideal
gas, which is proportional to $n^{2/3}$.
Note that the critical temperature always obeys 
$T_{c} \ll \hbar \omega / k_{B}$, which {\it a posteriori} shows that a 
single-band approximation for the dressed molecules is indeed appropriate.

{\it Conclusions and Discussion.} ---
In summary we have shown that
by using a red-detuned optical lattice with a wavelength of $806$~nm
the Bose-Fermi 
mixture of  $^{40}$K and $^{87}$Rb atoms can be
accurately described by a generalized Hubbard model.
Moreover, we have shown that the model contains a quantum phase transition
associated with the Bose-Einstein condensation of rubidium.
Interestingly, the presence of a Bose-Einstein condensate  induces
also a macroscopic coherence between the fermionic atoms and molecules, 
because of the specific form of the atom-molecule coupling near a Feshbach 
resonance. What is especially interesting is that such a coherence cannot be
obtained by solely making use of lasers in this case because 
it involves a quantum coherence between two different species.
It would, therefore, be very exciting to observe Rabi oscillations between 
fermionic atoms and molecules by an appropriate manipulation 
of the atomic Bose-Einstein condensate density. 

Using the known atomic physics of the Feshbach resonance to determine
the parameters in the generalized Hubbard hamiltonian
we have calculated the critical surface
for low filling fractions as a function of the applied magnetic field and 
temperature. Our analysis of the quantum phases of a resonantly-interacting 
Bose-Fermi mixture has been based on mean-field theory. In particular,
this implies that we have not considered the attractive finite-range
interaction between the fermionic atoms that can be mediated by density
fluctuations in the Bose-Einstein condensate \cite{Bijlsma,Heiselberg2000a}.
In principle this mechanism can lead to a BCS pairing between the fermionic 
atoms. However, for the spin-polarized mixture discussed here
this pairing must take place in 
a p-wave channel, which is expected to have a very small critical temperature
at small filling fractions. We, therefore, do not consider this interesting
possibility in detail here and leave that to future investigations.

This work is supported by the Stichting voor Fundamenteel Onderzoek der
Materie (FOM) and by the Nederlandse Organisatie voor
Wetenschaplijk Onderzoek (NWO).
D. van Oosten acknowledges funding through a Marie-Curie Excellence grant.

\bibliographystyle{apsrev}
\end{document}